\documentclass[conference,10pt]{IEEEtran}
\IEEEoverridecommandlockouts
\usepackage[nocompress]{cite}
\usepackage{amsmath,amssymb,amsfonts}
\usepackage{graphicx}
\usepackage{textcomp}
\usepackage{xcolor}
\usepackage{multirow}
\usepackage{subfig}
\usepackage{hyperref}
\usepackage[linesnumbered,ruled,vlined]{algorithm2e}
\usepackage[compatibility=false]{caption}

\def\BibTeX{{\rm B\kern-.05em{\sc i\kern-.025em b}\kern-.08em
    T\kern-.1667em\lower.7ex\hbox{E}\kern-.125emX}}
\begin{document}

\title{Study of AP Association and Users and Power Allocation for Cell-Free Massive MIMO Systems
 \thanks{The work of  S. Mohammadzadeh and K. Cumanan was supported by the UK Engineering and Physical Sciences Research Council (EPSRC) under grant number EP/X01309X/1.}
}


\author{S. Mohammadzadeh\textsuperscript{*},
       S. Mashdour\textsuperscript{\dag},
       R. C. de Lamare\textsuperscript{* \dag},
    K. Cumanan\textsuperscript{*},
    C.~Li\textsuperscript{*}
          \\
    \IEEEauthorblockA{\textsuperscript{*}School of Physics, Engineering and Technology - University of York,
York, United Kingdom \\
\textsuperscript{\dag}Centre for Telecommunication Studies, PUC-Rio, Brazil \\
 $ \rm  saeed.mohammadzadeh, rodrigo.delamare, kanapathippillai.cumanan cl2215@york.ac.uk\textsuperscript{*} $,
 $ \rm smashdour@gmail.com{\dag}$. 
}
    }

\maketitle
\begin{abstract}
This paper introduces an access point-user (AP-UE) association strategy combined with pilot power allocation to mitigate multiuser interference and enhance spectral efficiency (SE) in clustered cell-free massive MIMO (CCF-mMIMO) networks. We propose a dynamic channel-based clustering method that groups APs according to their channel correlation, ensuring users are associated with APs exhibiting similar channel characteristics. The proposed approach exploits hierarchical clustering, enabling flexible cluster sizing to improve interference management and overall SE. Moreover, we present a power control (PC) technique that is based on a weighted sum-rate maximization (WSRM) algorithm to ensure consistent service quality across users. Numerical results demonstrate that the proposed method achieves superior SE and robust performance in high-density multi-user environments as compared to competing approaches. 
\end{abstract}


\section{Introduction}

Cell-free massive multiple-input multiple-output (CF-mMIMO) systems have emerged as a key technology for future wireless networks, offering seamless connectivity and high service quality to users \cite{ngo2017cell} beyond centralized or co-located mMIMO systems \cite{mmimo,wence}. Unlike conventional cellular architectures \cite{jpba}, CF-mMIMO eliminates cell boundaries by deploying a large number of distributed access points (APs) that jointly serve users across a wide area. These APs are connected to a central processing unit (CPU) through fronthaul links, which in turn communicate via a backhaul network \cite{demir2021foundations}. CF-mMIMO systems benefit from enhanced spectral and energy efficiency, favorable propagation, high macro-diversity, and robust resource allocation \cite{chong2024performance}.

Standard CF-mMIMO networks operate with all APs serving all users simultaneously \cite{bjornson2019making,rmmsecf}. However, as the number of APs and users grows, the required channel state information (CSI) measurements increase linearly, leading to increased signaling overhead and computational complexity \cite{bjornson2020scalable}. Since nearby APs mostly influence a user’s signal power due to path loss, fully cooperative AP-UE association is inefficient. To address these challenges, user-specific AP clustering methods such as the power-based selection method \cite{ngo2017total}, localized clustering based on coverage areas \cite{lin2019user,cesg}, structured pilot assignment-based method \cite{chen2020structured,cdidd,iddllr,oclidd}, game-theoretic approach \cite{wei2022user}, and distance-based or random selection algorithm \cite{sheikh2022capacity,rscf} have been introduced to enhance network scalability. Furthermore, sum-rate-based clustering has shown improvements in performance \cite{mashdour2024clustering}, while machine learning techniques, including hierarchical clustering \cite{yemini2019virtual} have gained traction for AP selection.

Another aspect of cell-free networks is power control (PC), which plays a key role in optimizing resource allocation by reducing interference, improving user quality of service (QoS), and minimizing energy consumption at the CPU \cite{zhao2020power,rra}. Efficient PC strategies help exploit spatial diversity, mitigate pilot contamination, and enhance system performance \cite{van2018large}. However, designing PC algorithms in large-scale CF-mMIMO networks is complex due to optimization challenges and signaling constraints. PC methods designed for cellular systems \cite{van2020power} are not directly applicable to cell-free networks, where APs estimate wireless channels based on uplink pilot signals.

Several studies have used optimization techniques to address power control (PC) in CF-mMIMO networks. In \cite{ngo2017cell}, a bisection search algorithm was employed to solve the max-min PC problem in the downlink, while \cite{mai2018pilot} proposed a pilot power control scheme to minimize channel estimation errors. Joint PC strategies have been explored for energy-constrained APs \cite{masoumi2018joint} and further optimized using first-order methods \cite{mai2020design}. User-centric PC techniques have incorporated convex approximation and geometric programming \cite{braga2021joint}, while other approaches leverage sequential convex optimization \cite{liu2022joint} and weighted minimum mean square error-based algorithms \cite{chakraborty2020efficient}. Recently, machine learning- driven PC solutions have been investigated to enhance power allocation efficiency, including deep neural networks \cite{salaun2022gnn}. Additionally, our previous work \cite{mohammadzadeh2025pilot} contributes to this field by exploring adaptive pilot and data power allocation strategies.

Motivated by these challenges, this work proposes a dynamic AP selection and pilot power allocation scheme to enhance spectral efficiency (SE) in uplink CCF-mMIMO systems. A novel AP clustering method is introduced based on channel correlation and hierarchical clustering, ensuring optimal user association while improving interference mitigation and beamforming gains. The proposed framework enables scalable clustering by adjusting the group of APs without a full network reclustering, reducing CSI overhead and computational complexity. Unlike distance-based clustering methods, which are based on a fixed distance between APs and the users, the correlation-based approach enhances SE, considering the channels between the APs and users while effectively balancing backhaul communication requirements. Additionally, pilot power allocation strategies are developed based on a weighted sum-rate maximization (WSRM) technique that ensures a uniform QoS that dynamically assigns pilot power based on channel conditions to minimize interference. Numerical results confirm that the proposed framework significantly improves SE in high-density CF-mMIMO networks compared to existing clustering and PC techniques.\\
\textit{Notations:} We represent vectors using bold lowercase letters and matrices using bold uppercase letters. $\mathbb{R}$ represents real numbers. A circular symmetric complex Gaussian matrix with covariance $\mathbf{X}$ is denoted by $\mathcal{CN}(0,\mathbf{X})$ and $\mathbb{E} \{\cdot\}$ stands for the statistical expectation of random variables $(\cdot)^\mathrm{T}$, $(\cdot)^\mathrm{*}$. $(\cdot)^{-1}$ and $(\cdot)^\mathrm{H}$ denote transpose, conjugate, inverse, and conjugate-transpose, respectively. The Euclidean norm and the absolute value are represented by $\|\cdot \|$ and $|\cdot |$, respectively. We use the superscripts $^\mathrm{p}$ and $^\mathrm{d}$ to represent the variables or parameters associated with the pilot and data. 

\section{System Model}

We consider a CF-mMIMO system with $M$ randomly distributed APs serving $K$ single-antenna UEs, where $K \ll M$. Each AP is equipped with $L$ antennas and is connected to a CPU via fronthaul links. The system operates in time-division duplex (TDD) mode, and we assume perfect synchronization between all APs and users. The system is modeled in a block-fading channel, with the time and frequency plane divided into coherence blocks, $\tau_c$. Within these blocks, $\tau_p$ represents the uplink pilot symbols. The channel $\mathbf{g}_{mk}$ between the $m^\text{th}$ AP and the $k^\text{th}$ user is modeled as a combination of large-scale fading and small-scale fading, where $\mathbf{g}_{mk} = \sqrt{\beta_{mk}} \mathbf{h}_{mk}$. Here, $\beta_{mk}$ represents the large-scale fading, and $\mathbf{h}_{mk} \in \mathcal{CN}(0, \mathbf{I}_M)$ represents the small-scale fading coefficients. We assume that the $m^\text{th}$ AP has perfect knowledge of the local statistical CSI for all connected links, though this information can be estimated using standard methods if not perfectly known \cite{sanguinetti2013random}.

\subsection{Uplink Training and Channel Estimation}

In this work, we consider a clustered cell-free massive MIMO (CCF-mMIMO) network, where a selected subset of APs serves each user, denoted as \(\mathcal{C}_k\), as described in section \ref{AP clustering}. We assume the availability of \(\tau_p\) mutually orthogonal pilot sequences, represented as \(\sqrt{\tau_p} \boldsymbol{\psi} \in \mathbb{C}^{\tau_p \times 1}\), where \(\|\boldsymbol{\psi}\|^2 = 1\). The APs perform channel estimation using these pilot signals. For the \(k\)th user, the assigned pilot sequence is \(\sqrt{\tau_p} \boldsymbol{\psi}_k\). This follows a similar strategy to those used for co-located mMIMO systems \cite{spa,mfsic,mbdf,did,bfidd,msgamp}. The received pilot signal at the $m^\text{th}$ AP is given by:  
\begin{equation} \label{RPS}
\mathbf{Y}_{m}^{\mathrm{p}} = \sum _{k=1}^{K} \sqrt{\tau_p q_{k}^{\mathrm{p}}} \mathbf{g}_{mk} { \boldsymbol{\psi }}^{\mathrm{H}}_{k} + \mathbf{Z}_{m}^{\mathrm{p}},  
\end{equation}  
where \(q_k^\mathrm{p}\) denotes the pilot transmission power of the $k^\text{th}$ user, and \(\mathbf{Z}_{m}^{\mathrm{p}} \in \mathbb{C}^{L \times \tau_p}\) represents the additive white Gaussian noise matrix with independent and identically distributed elements following \(\mathcal{CN}(0, \sigma^2 \mathbf{I}_L)\), where \(\mathbf{I}_L\) is the identity matrix. To mitigate interference from other users, the received pilot matrix \(\mathbf{Y}_{m}^{\mathrm{p}}\) is projected onto the pilot sequence \(\boldsymbol{\psi}_k^\mathrm{H}\), yielding  $\mathbf{y}_{mk}^\mathrm{p} = \mathbf{Y}_{m}^{\mathrm{p}} \boldsymbol{\psi}_k$. We employ the minimum mean square error (MMSE) estimator to estimate the channel \(\mathbf{g}_{mk}\), given by:  
\begin{equation}
\hat{\mathbf{g}}_{mk} = \frac{\mathbb{E} \{ \mathbf{y}_{mk}^\mathrm{p} \mathbf{g}_{mk}^* \}}{\mathbb{E} \{ \mathbf{y}_{mk}^\mathrm{p} (\mathbf{y}_{mk}^\mathrm{p})^\mathrm{H} \}} \mathbf{y}_{mk}^\mathrm{p} = \sqrt{\tau_p q_k^{\mathrm{p}}} \beta_{mk} \zeta_{mk}^{-1} \mathbf{y}_{mk}^\mathrm{p},  
\end{equation}  
where \(\hat{\mathbf{g}}_{mk}\) represents the estimated channel coefficient, and \(\zeta_{mk}\) is given by:  
\begin{equation}  \label{gamalu}
   \zeta_{mk} = \mathbb{E} \{ \mathbf{y}_{mk}^\mathrm{p} (\mathbf{y}_{mk}^\mathrm{p})^\mathrm{H} \} =  \tau_p \sum \limits _{j=1}^K q_{j}^{\mathrm {p}}\beta _{mj}| { \boldsymbol {\psi }}^{\mathrm {H}}_{k} \boldsymbol {\psi }_{j}|^{2} + \sigma^2.  
\end{equation}  
To ensure efficient network scalability, we define an association matrix \(\mathbf{B} \in \mathbb{R}^{M \times K}\), where \(\mathbf{B}_{mk} = 1\) indicates that AP \(m\) serves the $k^\text{th}$ user, and \(\mathbf{B}_{mk} = 0\) otherwise. The set of APs serving $k^\text{th}$ user is then defined as \(\mathcal{C}_k = \{ m : \mathbf{B}_{mk} = 1, m \in \{1, \cdots, M\} \}\).

\subsection{Uplink Data Transmission and Achievable SE}
During uplink data transmission, all users transmit their data to the APs using the same time-frequency resources. The received signal at the $m^\text{th}$ AP is given by
\begin{equation} 
\mathbf {y}_m^{\mathrm {d}} = \sum _{k=1}^{K} \sqrt{{q_{k}^\mathrm {d}}} \mathbf {g}_{mk}s_{k} + \mathbf {z}_{m}^{\mathrm {d}}, 
\end{equation}
where $q_k^\mathrm{d}$ is the transmit data power of the $k^\text{th}$ user, $s_k$ is the data symbol for user $k$ (with $\mathbb{E}[|s_k|^2] = 1$), and $\mathbf{z}_m^{\mathrm{d}} \sim \mathcal{CN}(0, \sigma^2 \mathbf{I}_M)$ is the noise at the receiver. The $m^\text{th}$ AP employs maximum ratio combining (MRC), multiplying the received signal by the Hermitian transpose of the channel estimate $\hat{\mathbf{g}}_{mk}^\mathrm{H}$, and sends this to the CPU for further processing. The received signal for user $k$ is $y_k = \sum_{m \in \mathcal{C}_k} \hat { \mathbf {g}}_{mk}^\mathrm{H} \mathbf {y}_m^{\mathrm {d}}$. By applying the use-and-then-forget technique \cite{marzetta2016fundamentals}, combined received signal can be written as follows
\begin{align} \label{ru}
y _ { k } = \text{DS}_k s_k + \text{BU}_k s_k + \sum _ { j \neq k } ^ { K } \text{IUI}_{kj}  +  \text{TN}_k,
\end{align}
where $\text{BU}_k =  \sqrt { q_{k} ^ \mathrm{ d } } \sum _ {m \in \mathcal{C}_k } \mathbf{g} _ {mk} \hat { \mathbf{g} }_{mk} ^ \mathrm{H} - \mathrm { DS }_k $, and $\text{DS}_k = \mathbb { E } \{{ \sum _ { m \in \mathcal{C}_k }  \sqrt {q_k^\mathrm{d}} \mathbf{g}_{mk} \hat {\mathbf{g}}_{mk}^\mathrm{H}} \}$, $\text{TN}_k =\sum_{m \in \mathcal{C}_k} \mathbf{n}_{m}^\mathrm{d} \hat{\mathbf{g}}_{mk}^\mathrm{H} $ , $\text{IUI}_{kj} = \sqrt { q_j ^ \mathrm{d} } \sum _ { m \in \mathcal{C}_k } \mathbf{g}_{mj } \hat { \mathbf{g} }_{mk} ^ \mathrm{H} s_j$. The SINR of the received signal in \eqref{ru} is defined by considering the worst-case scenario with uncorrelated Gaussian noise, as follows:
\begin{align} \label{fin SINR}
&\mathrm{SINR}_k \nonumber \\& = \dfrac{L \tau q_k^\mathrm{d} q_k^\mathrm{p}  \sum_{m \in \mathcal{C}_k}\beta_{mk}^2 {\zeta}_{mk}^{-1} }{ \sum_{m=1}^M \big( L \tau \sum _{\substack{j=1\\ j\ne k}}^{K} p_j^\mathrm{d} p_j^\mathrm{p} \beta_{mj}^2 {\zeta}_{mk}^{-1} +  \sum_{j=1}^K p_j^\mathrm{d}  \beta_{mj}  + \sigma^2\big)}.
\end{align}
and the achievable uplink rate of the $k^\text{th}$ user is given by
\begin{equation} \label{Original Rate}
\text{SE}_{k} = (1 - \frac{\tau}{\tau _{\text {c}}})\log _{2}\left ({1 +  \mathrm{SINR}_{k}}\right),
\end{equation}
\section{Proposed AP clustering and Pilot power allocation }
In this section, we introduce a novel, efficient dynamic AP clustering method based on channel correlation, ensuring that the APs serve users with the most similar channels. Moreover, we propose a method to optimize pilot power allocation to improve each user's SINR during the training phase. This approach focuses on the WSRM problem between an AP and users while ensuring reliable communication.
\begin{figure}[!]
	\centering
\includegraphics[width=2.3in,height=1.7in]{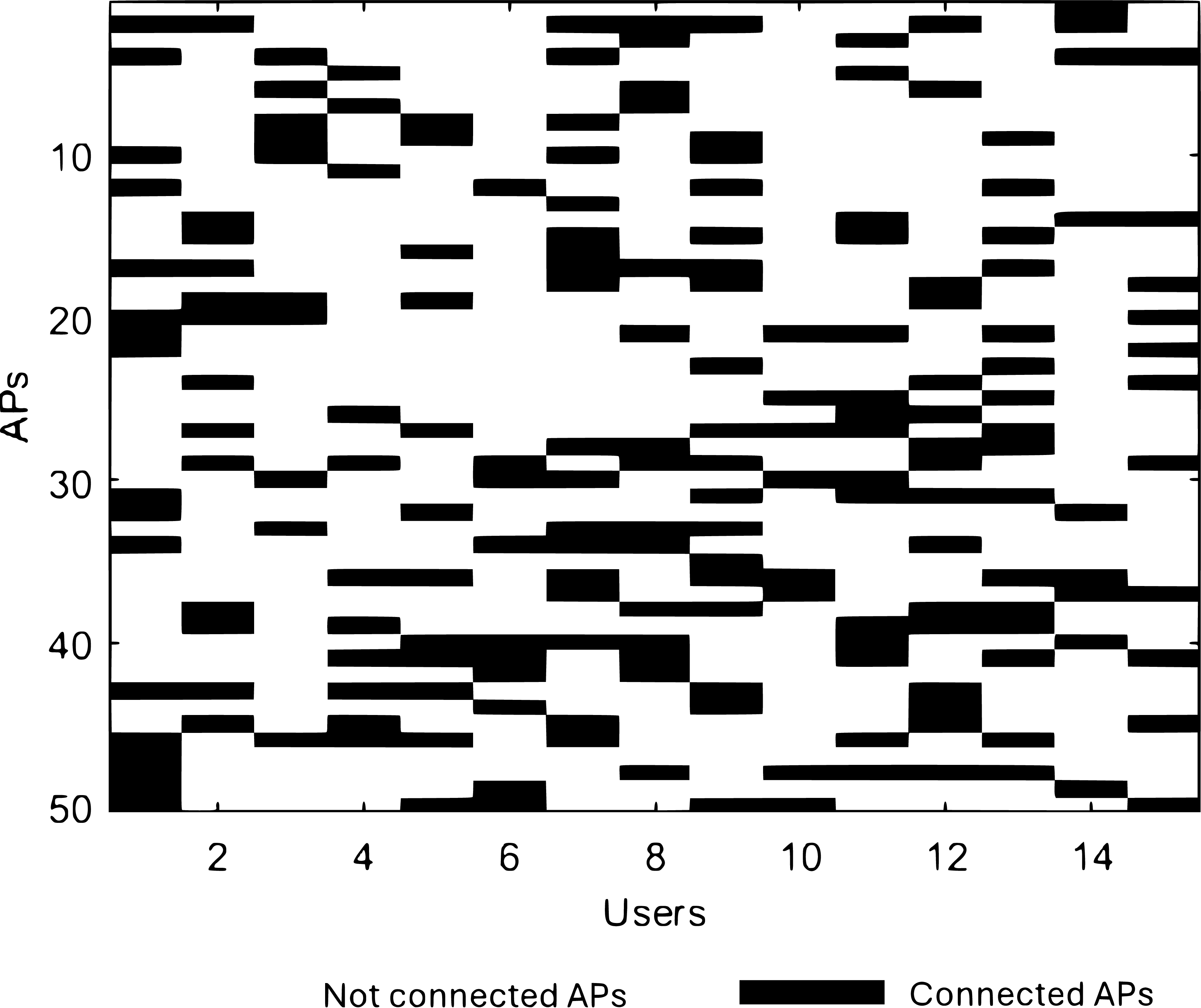}
	\caption{An example of allocated APs for each user}
 \label{AP distribution}
\end{figure} 
 
\subsection{Access points clustering} \label{AP clustering}
The approach involves dynamic clustering APs based on the correlation between their channel responses with each user. This ensures that APs serving each user have closely aligned channel characteristics and can jointly optimize the user’s signal reception, leading to better connectivity and more effective interference management across the network. In the proposed clustering formation as shown in Fig.~\ref{AP distribution}, we aim to group a subset of the APs that provide similar channel conditions to the $k^\text{th}$ user dynamically. In this regard, the collective channel from all users to $m^\text{th}$ AP is defined as $\mathbf{g}_m = [\mathbf{g}_{m1}^\mathrm{T}, \mathbf{g}_{m2}^\mathrm{T}, \ldots, \mathbf{g}_{mk}^\mathrm{T}]^\mathrm{T} \in \mathbb{C}^{LK}$. Since we want to find the APs with the most similar channels that serve the $k^\text{th}$ user, the correlation coefficient between the channels of two APs, such as $m$ and $n$, can be expressed as:
\begin{align}
    \rho_{mn} = \frac{\mathbf{g}_m^\mathrm{H}\mathbf{g}_n}{ \| \mathbf{g}_m \| \| \mathbf{g}_n \|}, \quad \forall m,n \in \{ 1,2, \cdots, M\}
\end{align}
This evaluates the correlation between the channel vectors of different APs. APs with high correlation should be clustered together. To this end, we define a distance matrix $\mathbf{D} \in \mathbb{R}^{M \times M}$ and its entries given by
        \begin{align}
             D_{mn} = 1 - \rho_{mn}.
        \end{align}
\indent \textit{Note that in our method, the distance matrix does not represent the physical distances between pairs of APs}. Instead, it quantifies the difference in their correlation values, providing a measure of similarity between the APs based on their channel characteristics. Each entry in the matrix corresponds to the computed "distance" in terms of correlation rather than geographic separation. After that, we utilize hierarchical clustering to group APs based on their channel correlation. The clustering is based on the correlation coefficients $\rho_{mn}$ and the linkage function, which defines how to measure the distance between sets of data points. \\
\indent In hierarchical clustering, the process starts with each AP as an individual cluster, then iteratively merges the closest clusters based on the linkage function that computes the steps of this merging process. Then, the APs are clustered based on the threshold distance. Lastly, for each user, $k$, by sorting the large-scale fading coefficients between AP $m$ and all users in descending order, the algorithm selects the APs from the top cluster with respect to the $k^\text{th}$ user, and finally, the APs that serve user $k$ are defined. 
The proposed approach uses channel-based clustering to dynamically group APs with similar channel conditions, reducing inter-AP interference and improving spectral efficiency. Unlike distance-based methods, it adapts to real-time channel variations, serving each user through APs with the strongest links. This enhances performance, scalability, and backhaul efficiency without relying on complex SE metrics.

\subsection{Pilot power allocation}
The CF-mMIMO system operates in TDD mode, with channels estimated at each AP based on user pilot signals during the training phase. However, due to the non-orthogonality of pilot sequences, the channel estimates for a given user are corrupted by interference from pilots sent by other users, a phenomenon known as pilot contamination \cite{ngo2017cell}. This issue is particularly pronounced where the goal is to simultaneously serve a large number of users within the same time-frequency resources. As a result, mitigating pilot contamination is critical. To tackle pilot contamination, recent studies have introduced various algorithms aimed at mitigating its effects, especially in densely populated multi-user settings, to improve communication quality by reducing interference from pilot reuse \cite{ osawa2023overloaded}.\\
However, this work assumes a scenario with substantial pilot contamination resulting from random pilot assignment. This assumption allows for evaluating the proposed method’s performance under challenging, worst-case conditions. Thus, we propose a solution to optimize pilot power allocations to maximize each user's SINR during the training phase.

Given the objective of providing a better quality of service to all users, we formulate this as a WSRM problem \cite{weeraddana2012weighted}. The system dynamically adjusts pilot powers to maximize the weighted sum of users' data rates, subject to power constraints as follows
\vspace{-1.5em}
 \begin{align} \label{WSR}
P_1: \quad  & \max _{q_k^\mathrm{p}} \sum _{k=1}^K w_k \text{SE}_{k}(q_k^\mathrm{p}) \nonumber  \\ &\text{s.t}  \quad  \epsilon \leq  q_k^\mathrm{p}  \leq  P_{\text{max}} \quad \forall k ,
\end{align}
where $w_k$  is a weight associated with the $k^\text{th}$ user (representing user priority) and $P_{\text{max}}$ denotes the maximum transmit budget available for each user. In this paper, since we have assumed that users are treated equally (no priority given to any specific user), the weight vector can be defined as $w_k = \mathbf{1}_K$. It is clear from the optimization problem in \eqref{WSR} that the SE for each user $k$ depends on their SINR, which is a power allocation function. Therefore, the optimal power can be determined by solving the following maximization problem:
\begin{align}\label{WSR SINR}
 P_2: \quad   & \max _{q_k^\mathrm{p}} \quad \sum _{k=1}^K \text{SINR}_{k}(q_k^\mathrm{p})  \nonumber  \\ &\text{s.t}  \quad  \epsilon \leq  q_k^\mathrm{p}  \leq  P_{\text{max}} \quad \forall k,
\end{align}
It is seen that the problem \eqref{WSR SINR} is a non-convex optimization problem with constraint variable $q_k^\mathrm{p}$. To deal with this optimization problem, we apply the log-sum-exp (LSE) approximation \cite{boyd2004convex}. The function LSE is defined as
\begin{align} \label{LSE}
    \max(x_1, x_2, \cdots, x_n) \approx \lim_{\lambda \rightarrow \infty} (1/\lambda) \log \big( \sum _{i=1}^n e^{\lambda x_i} \big),
\end{align}
and is a convex function on $\mathbb{R}^n$. This function effectively approximates the maximum of a set of values when the terms $x_i$ differ significantly, as the largest value dominates it. This makes LSE useful in log-likelihood calculations and in replacing the max function in smooth optimization problems, allowing for gradient-based optimization methods. By applying \eqref{LSE} to \eqref{WSR SINR}, we can write
\begin{equation} \label{MAXSINR}
\mathop {\max }\limits_k {\text{SIN}}{{\text{R}}_k}(q_k^p) \cong ( 1/\lambda )\log \Big( {\sum\limits_{k = 1}^K {e^{\left( { \lambda \text{SINR}_k}(q_k^p) \right)}} } \Big),\lambda \to \infty 
\end{equation}
In this equation, \(\lambda\) is a parameter that controls the smoothness of the maximum function's LSE approximation. As \(\lambda \to \infty\), the expression converges to the true maximum of the \(\text{SINR}_k\) values, effectively approximating the max operator. For finite values of \(\lambda\), this approximation is smooth and differentiable, which is useful in optimization, but the larger \(\lambda\), the closer the result will be to the actual maximum \(\text{SINR}\) value across users. The optimization problem in \eqref{MAXSINR} can be written as 
\begin{align}
    \max _{q_k^\mathrm{p}} \quad X (q_k^\mathrm{p}),
\end{align}
where $X \triangleq ( 1/\lambda )\log \Big( {\sum\limits_{k = 1}^K {e^{\left( { \lambda \text{SINR}_k}(q_k^p) \right)}} } \Big)$.
To address this problem, we present an iterative gradient descent (GD) algorithm described in Algorithm 1, where we define
\begin{align}\label{gradSINR}
\nabla_{q_k^\mathrm{p}}& \mathrm{SINR}_k(q_k^p) \nonumber \\ = &\frac{\text{De}_k(q_k^\mathrm{p})\cdot  \nabla \text{No}_k(q_k^\mathrm{p}) - \text{No}_k(q_k^\mathrm{p}) \cdot \nabla \text{De}_k(q_k^\mathrm{p})}{\big(\text{De}_k(q_k^\mathrm{p})\big)^2},
\end{align}
and $\text{No}_k(q_k^\mathrm{p}) \triangleq L \tau q_k^\mathrm{d} q_k^\mathrm{p} \sum_{m \in \mathcal{C}_k} \beta_{mk}^2 \zeta_{mk}^{-1}$ and $\text{De}_k (q_k^\mathrm{p}) \triangleq \sum_{m \in \mathcal{C}_k} \Big( L \tau \sum_{\substack{j=1\\ j \neq k}}^K p_j^\mathrm{d} p_j^\mathrm{p} \beta_{mj}^2 \zeta_{mk}^{-1} + \sum_{j=1}^K p_j^\mathrm{d} \beta_{mj} + \sigma^2 \Big)$.
The key approach involves alternately optimizing \( q_k^\mathrm{p} \) while keeping the other variable constant.

\begin{algorithm}[t]
\caption{GD algorithm to calculate pilot power }
\label{alg_1}
\SetAlgoLined
 Calculate $\nabla_{\mathbf{q}_u^\mathrm{p}} \mathrm{SINR}_k $ using \eqref{gradSINR}\;
  Set $\mathbf{z}_0= \nabla_{\mathbf{q}_u^\mathrm{p}} \mathrm{SINR}_k $ and $\mathbf{d}_0=-\mathbf{z}_0$\;  
   Set $t \gets 0$\;  
  \While{$\Vert \nabla_{\mathbf{q}_k^\mathrm{p}} \mathrm{SINR}_k ^{(t+1)} \Vert > \kappa $}{
 \qquad Determine the step-size $\alpha_t=\frac{- \mathbf{z}_t  \mathbf{d}_t}{\mathbf{d}^{T}_t \mathbf{d}_{t}} $\; 
 \qquad $\mathbf{q}_u^{\mathrm{p}_{(t+1)}} = \mathbf{q}_u^{\mathrm{p}_{(t)}} + \alpha_t \mathbf{d}_t$\;
 \qquad Compute $ \mathbf{z}_{(t+1)} = \nabla_{\mathbf{q}_u^\mathrm{p}} \mathrm{SINR}_k ^{(t+1)} $\;
  \qquad Determine  $\zeta_t={\mathbf{z}_{t+1}^\mathrm{T} (\mathbf{z}_{t+1}-\mathbf{z}_{t})} / {  \parallel \mathbf{z}_{t} \parallel^2}$\;
 \qquad Set $\mathbf{d}_{t+1}=-\mathbf{z}_{t+1}+\zeta_t\mathbf{d}_t$\;
  \qquad Set $t \gets t+1$\;
      }
 $\textbf{Output:} \quad \hat{\mathbf{q}}_k^\mathrm{p}$
\end{algorithm}
\section{Simulation Results}
\begin{figure}[ht]
	\centering
\includegraphics[width=3.0in,height=1.9in]{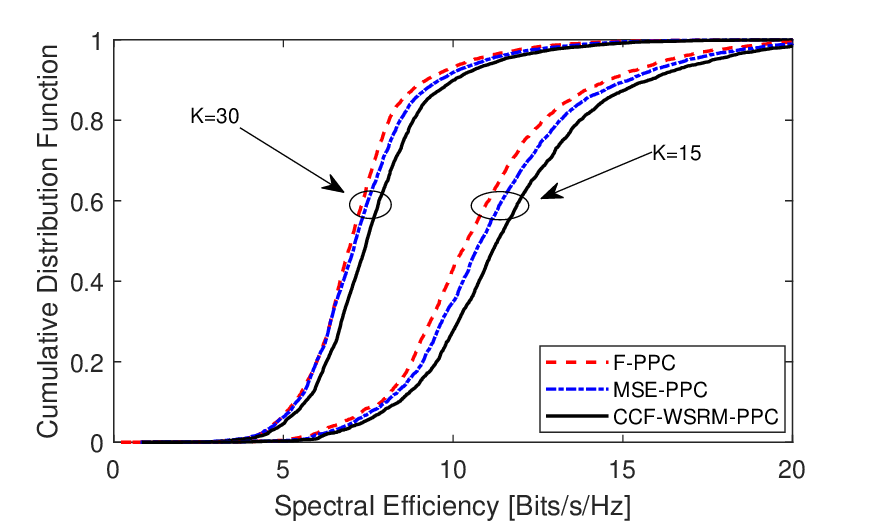}
	\caption{SE per UE for $\tau_p=10$ and $L=1$}
 \label{ALL method}
\end{figure}

The simulation framework and parameter settings align closely with those presented in \cite{ngo2017cell}. Specifically, 50 APs with $L$ antennas and $K$ users are randomly distributed within a square area of $0.5 \times 0.5$ Km, employing a wrap-around technique to emulate an infinitely extended network. The large-scale fading coefficient is modeled by incorporating both path loss and uncorrelated log-normal shadowing. The path loss is characterized using a three-slope model. The numerical results are averaged over 300 independent random realizations of AP and user placements to ensure statistical reliability. The proposed CCF-WSRM-PPC algorithm is compared to he full pilot power control (F-PPC) method in \cite{ngo2017cell} and the mean-square error-based pilot power control (MSE-PPC) method in \cite{mai2018pilot}.
Fig.~\ref{ALL method} presents the cumulative distribution function versus spectral efficiency for different power pilot control methods in scenarios with $K = 15$ and $K = 30$ users when each AP is equipped with $L = 1$ antenna. Increasing the number of users raises interference, lowering SE. For 30 users, F-PPC yields slightly lower SE than other methods. However, the MSE-PPC outperforms the F-PPC, while the proposed WSRM-PPC performs best. When the number of users reaches 15, notably, WSRM-PPC maintains superior SE across most levels, highlighting its robustness in densely populated user environments.

\begin{figure}[ht]
\centering
\subfloat[$\tau_p = 5$]{\label{N100tau5} \includegraphics[height=1.8in]{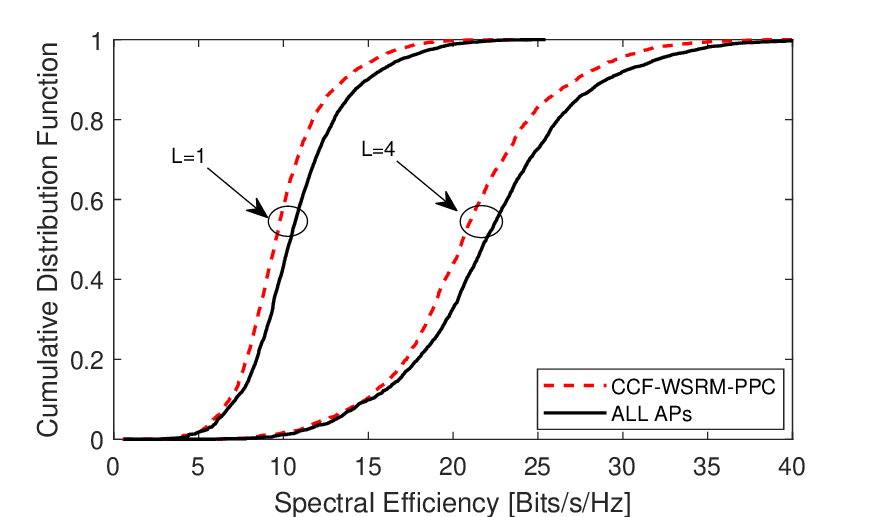}} 
\qquad
\subfloat[$\tau_p = 15 $]{\label{N100tau10} \includegraphics[height=1.8in]{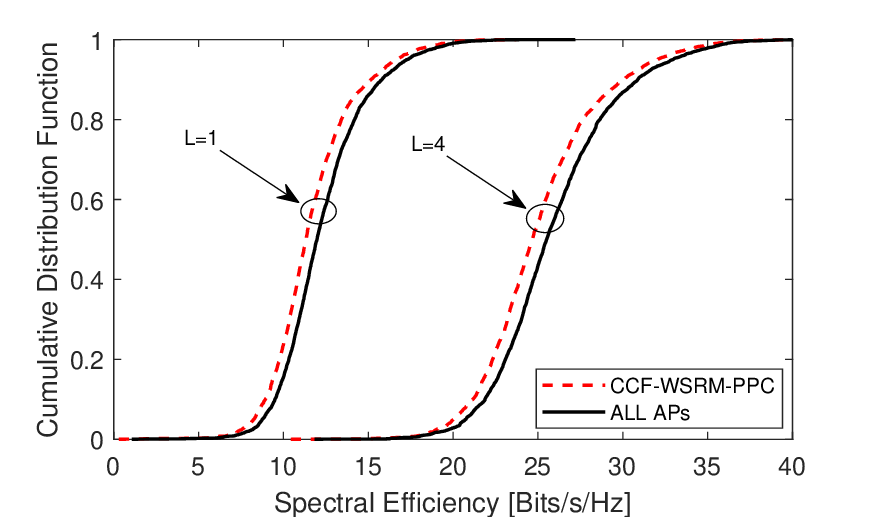}} 
\vspace{-0.25em}
\caption{SE per UE of the proposed method for a different number of antennas ($L$) and $K=15$.  a) random pilots $\tau_p=5$, b) orthogonal pilots $\tau_p=15$}
\label{All versus cluster}
\end{figure}

\indent Fig.~\ref{All versus cluster} presents the performance of the proposed CCF-WSRM-PPC method in the scenario in which all APs and the clustered APs schemes are compared when the pilot sequences are considered as $\tau_p =5$ and $\tau_p=15$.  It is seen that when the number of antennas is increased, the performance loss remains minimal for both scenarios with fewer pilot sequence numbers than the number of users. This loss mainly arises from increasing the number of antennas that serve each user and employing the proposed scalable clustering algorithm. However, since the impact on performance is small, the trade-
off for scalability is also minimal, and the algorithm maintains overall solid performance.

\section{Conclusion}
This work proposes a joint AP selection and pilot power allocation scheme for uplink CF-mMIMO systems. By using channel-based hierarchical clustering, users are associated with APs having strong, correlated channels to reduce inter-AP interference. The flexible clustering adapts to real-time conditions, while a WSRM-based power control improves SINR during training, enhancing overall spectral efficiency.

\bibliographystyle{IEEEtran}
\bibliography{refs}

\end{document}